\documentclass[12pt,preprint]{aastex}

\shorttitle{$K_S$ -band Selected High Redshift Galaxy Populations in the $AKARI$ NEP Field}
\shortauthors{Imai et al.}

\begin{document}

%% LaTeX will automatically break titles if they run longer than
%% one line. However, you may use \\ to force a line break if
%% you desire.

\title{Number Density Evolution of \\
         $K_S$ -band Selected High Redshift Galaxy Populations in \\
         the $AKARI$ North Ecliptic Pole Field}

%% Use \author, \affil, and the \and command to format
%% author and affiliation information.
%% Note that \email has replaced the old \authoremail command
%% from AASTeX v4.0. You can use \email to mark an email address
%% anywhere in the paper, not just in the front matter.
%% As in the title, use \\ to force line breaks.

\author{Koji Imai}
\affil{TOME R\&D Inc. Kawasaki, Kanagawa 213 0012, Japan}
\email{koji@ir.isas.jaxa.jp}

\author{Chris P. Pearson\altaffilmark{1}}
\affil{Space Science and Technology Department, CCLRC Rutherford Appleton Laboratory, Chilton, Didcot, Oxfordshire OX11 0QX, United Kingdom}

\author{Hideo Matsuhara, Takehiko Wada, Shinki Oyabu, Toshinobu Takagi, Naofumi Fujishiro}
\affil{Institute of Space and Astronautical Science, Japan Aerospace Exploration Agency, Sagamihara, Kanagawa 229 8510, Japan}

\and

\author{Hitoshi Hanami}
\affil{Iwate University, Morioka, Iwate 020 8550, Japan}

%% Notice that each of these authors has alternate affiliations, which
%% are identified by the \altaffilmark after each name.  Specify alternate
%% affiliation information with \altaffiltext, with one command per each
%% affiliation.

\altaffiltext{1}{Department of Physics, University of Lethbridge, 4401 University Drive, Lethbridge, Alberta T1J 1B1, Canada}

%% Mark off your abstract in the ``abstract'' environment. In the manuscript
%% style, abstract will output a Received/Accepted line after the
%% title and affiliation information. No date will appear since the author
%% does not have this information. The dates will be filled in by the
%% editorial office after submission.

\begin{abstract}
We present the number counts of $K_S$ -band selected high redshift galaxy
populations such as extremely red objects (EROs), $B$-, $z$- \& $K$ -band
selected galaxies (BzKs) and distant red galaxies (DRGs) in the $AKARI$ NEP field.
These high redshift galaxy samples are extracted from a multicolor catalog
combining optical data from Suprime-Cam on the 8.2 m Subaru telescope with
near-infrared data from the Florida Multi-object Imaging Near-IR Grism
Observational Spectrometer on the Kitt Peak National Observatory 2.1 m
telescope over 540 arcmin$^2$ in the NEP region
field. The final catalogue contains 308 EROs ($K_S<19.0$; 54\% are dusty star-forming
EROs and the rest are passive old EROs), 137 star-forming BzKs and 38
passive old BzKs ($K_S<19.0$) and 64 DRGs ($K_S<18.6$). We also produce individual component source counts for both the  dusty star-forming and passive populations.

We compare the observed number counts of the high redshift passively
evolving galaxy population with a backward pure luminosity evolution (PLE)
model allowing different degrees of number density evolution.
We find that the PLE model without density evolution fails to explain the observed counts
at faint magnitudes, while the model incorporating  negative density
evolution is consistent with the observed counts of the passively evolving
population.

We also compare our observed counts of dusty star-forming EROs with a
phenomenological evolutionary model postulating that the near-infrared EROs
can be explained by the source densities of the far-infrared --
submillimetre populations. Our model predicts that the dusty ERO source
counts can be explained assuming a 25 percent contribution of submillimetre
star-forming galaxies with the majority of brighter $K_S$ -band detected
dusty EROs having luminous (rather than HR10 type ultra-luminous)
submillimetre counterparts. We propose that the fainter  $K_{S}>$19.5 population is
dominated by the sub-millijansky submillimetre population. We also
predict a turnover in in dusty ERO counts around 19$<K_{S}<$20.
\end{abstract}

%% Keywords should appear after the \end{abstract} command. The uncommented
%% example has been keyed in ApJ style. See the instructions to authors
%% for the journal to which you are submitting your paper to determine
%% what keyword punctuation is appropriate.


\section{Introduction}
A quantification of the number density evolution of each galaxy population in successive epochs
would provide  stringent constraints on  galaxy formation models.
This decade has seen enormous progress in this field, particularly in the $K$ -band selected high redshift galaxy populations
such as the extremely red objects \citep[EROs, e.g.][sometimes also referred to as extremely red galaxies;
ERGs]{els88}, $B$-, $z$- and $K$ -band selected galaxies \citep[BzKs] {dad04} and distant red galaxies
\citep[DRGs] {van03,fra03}.

EROs are selected to have very red optical-to-near infrared colors such as $R-K>5-7$, $I-K>4-6$ (Vega magnitudes).
Spectroscopic studies have shown that observed EROs can be classified as either dusty star-forming galaxies,
old passive galaxies or AGN \citep[e.g.] {spi97,dey99,cim99,soi99,liu00}.
\citet{poz00} have shown that the old stellar population is expected to have a strong spectral break
between 3000 and 5000 \AA, while the effect of dust reddening in a younger population would produce a smoother spectra.
This effectively segregates the ERO population into dusty star-forming and passive old galaxies
(hereafter dERO and pERO, respectively) at $1<z<2$ in the $RJK$ or $IJK$ colour diagram.
Mannucci et al. (2002) have shown that the relative abundances of dEROs and pEROs is roughly equal using a sample of 57 EROs
($K'<20$ and $R-K'>5.3$) based on the classification method of \citet{poz00}
(Their sample was classified as 37\% elliptical, 37\% star-forming, 9\% stars with 17\% remaining unclassified).
Subsequent spectroscopy, morphological and X-ray observations have shown that the comoving space densities of
elliptical-like and disk-like EROs are comparable over the magnitude range $K_S<20$ \citep{cim02,mou04}
and that only a small fraction (5\%-10\%) of EROs host AGN \citep{roc03}.
The additional properties of strong clustering \citep{dad00,mcc01} and the relation of dEROs to (ultra)
luminous IR-galaxies has also been examined \citep{elb03}. Indeed dEROs have been detected in the infrared with both the $ISO$ observatory \citep{elb02}  and the $Spitzer$ space telescope \citep{wil04},with more than 12$\%$ of the 8--24$\mu$m sources being classified as dEROs in $Spitzer$ observations of the Lockman Hole.
%  12$\%$ of the 24$\mu$m sources being classified as dEROs comes from 22$\%$ of 60$\%$
dEROs have also been detected at longer submillimetre wavelengths by the Submillimetre Common User Bolometer Array (SCUBA) on the James Clerk Maxwell Telescope in Hawaii  \citep{frayer03,frayer04}. If dEROs are detected in significant numbers at infrared-submillimetre wavelengths then they may also be significant contributors to the dusty cosmic star-formation rate in the Universe.

Recently, \citet{dad04} have suggested a new criterion to isolate star-forming and passive evolving galaxies at $1.4<z<2.5$ in
the color-color parameter space based on $B$-, $z$- and $K$ -band photometry.
Previous studies of these BzK selected sources have shown that the star-forming galaxies (hereafter sBzKs) have star formation rates
of $\sim100 $M$_{\odot}$ yr$^{-1}$ and stellar masses of $10^{10-11} $M$_{\odot}$,
whilst the passive evolving galaxies (hereafter pBzKs) are found to have stellar masses of $>10^{11} $M$_{\odot}$
\citep{dad04,dad05,kon06,hay07}.
Both pBzKs and sBzKs are also found to be strongly clustered \citep{kon06,hay07}.

Distant Red Galaxies (DRG) are identified by their observed red near-infrared colors, $J-K>2.3$. 
\citet{fra03} argue that the DRG color criterion is designed to be sensitive to those galaxies with a large 4000 \AA 
$ $ break and produces a sample that is mainly populated by galaxies at $z>2$.
Subsequent analysis has concluded that DRGs have high star formation rates of $100-400$ M$_{\odot}$ yr$^{-1}$,
high stellar masses of $(1-5)\times  10^{11} $M$_{\odot}$ and are actively forming stars at $z\sim2-3.5$
\citep[e.g.] {rub04,van04,knu05,pap06}.

The contamination by galaxies at lower redshifts of the high-z populations selected via the DRG colour criterion has also been discussed in recent papers \citep{gra062,pap06} and
it has been shown that the high-z DRG population are more luminous and exhibit stronger clustering than that of the low-z population that is also selected with the DRG colour criterion.
From these studies, these high redshift galaxy populations are inferred to be the progenitors of the present-day early type galaxies.

In this work, we focus on the number counts of such $K$- band selected high redshift galaxy populations.
Based on a very deep sample ($K_S<22$) of 198 EROs from the ESO/GOODS data in the CDF-S,
\citet{roc03} show that the ERO number counts exhibit a change of slope flattening at $K>19.5$ from $d(\log N)/dm=0.59$ to 0.16.
Recent results of a deep survey with the Subaru/MOIRCS instrument in the GOODS-N field, \citet{kaj07} shows
that the number counts of DRGs also turn over at $K\sim22$ \citep[see also] {fou06,gra061}.
Although the number counts of sBzKs increase in number sharply toward fainter magnitudes,
those of the pBzKs also clearly turn over around $K\sim19$ \citep{kon06,lan07}.
This indicates that these  number counts are directly sampling the luminosity function (LF) at a specific epoch.
In this scenario, the corresponding absolute magnitude at the flattening of the number counts will be close to the value of
M$_{K}^*$ determined for the $K$ -band LF over the same redshift range \citep {kas03,poz03,sar06}.
Furthermore the trend of a late-type galaxies to posess a steeper slope of the LF \citep{fol99,mad02,nak03,dah05} is also
consistent with that of the sBzK counts exhibiting steeper slope at the faint end.
%In addition, recent a study of the $K$- band galaxy counts also show a existence of `bump' in the magnitude range $18.0<K_S<19.5$ as well as clear evidence of the slope change in the $J$ -band \citep{ima07}.

Note that some studies have claimed that the ERO counts are lower than those predicted by pure luminosity evolution (PLE) models.
\citet{roc02} derived the number counts of EROs using a sample ($K<21$) of 158 EROs in the ELAIS N2 field
and reported lower counts than models in which all E/S0 galaxies evolve according to PLE.
On the other hand, conflicting results have also been found from brighter samples of EROs.
\citet{vai04} show that the cumulative number counts of EROs ($K<17.5$) over a 2850 arcmin$^2$ area within the ELAIS field
\citep{row04} are well fited by PLE models.
However, several results implying a lower density of early-type galaxies at $z>1$ have been reported from morphological studies.
For instance, \citet{kaj01} reported that the comoving density of $M_V<-20$ galaxies decreases at $z>1$ in
each morphological category based on the results of the HST WFPC2/NICMOS archival data of the HDF-N.
In particular, they showed that the number density of the bulge-dominated galaxies conspicuously decreases to
1/5 of the local value between the $0.5<z<1.0$ and $1.0<z<1.5$ bins.
Other morphological studies have also revealed an apparent lack of massive ellipticals at redshifts greater than $z\sim1.5$ \citep[e.g.] {fra98,rod01}.

Hierarchical models of galaxy formation predict a significant decline in the comoving density of ellipticals with redshift,
as they should form through merging at $z<2$ \citep[e.g.] {whi78,whi91,kau96,bau96,som99,col00}
and a measure of a decline in their comoving density would provide valuable proof of these models.
It is therefore important to study the counts of these $K$ -band selected high redshift galaxy populations
because these samples will provide us with the clues for the formation and evolution of galaxies.

We describe our observations in the $AKARI$ north ecliptic pole (NEP) field in \S \ref{sec:nep_field}.
We then present our sample of the $K$ -band selected high redshift galaxy populations in \S
\ref{sec:Sample Selections of EROs, BzKs and DRGs}.
These observed number counts are presented in \S \ref{EROs_BzKs_and_DRGs_counts}.
In \S \ref{sec:discussion}, we discuss the relationship and trend of the ERO, BzK and DRG count slopes, 
the results of comparing with galaxy count models and implications from other observational studies.
Finally, the summary of this research is given in \S \ref{sec:summary}. 
Throughout this paper a concordance cosmology of H$_0=70$ kms$^{-1}$Mpc$^{-1}$,
$\Omega_{m}=0.3$ and $\Omega_{\Lambda}=0.7$ is assumed in agreement with the results for Type Ia supernovae
\citep{rie98,per99} and WMAP \citep{spe03}.
Unless otherwise stated, we use the $Vega$ magnitude system.

\section{$AKARI$ NEP field}
\label{sec:nep_field}
The Japan Aerospace Exploration Agency (JAXA) launched the infrared astronomy satellite 
$AKARI$ \citep[formerly $ASTRO$-$F$] {mur04} on February 22nd 2006 (Japan Standard Time).

Due to the fact that the orbit of $AKARI$ is Sun-synchronous polar, $AKARI$ can often obtain deep exposures at the ecliptic poles.
Such a deep survey program is currently under way in the $AKARI$ NEP field \citep{mat06}.
A portion of the field has also been covered by deep $B$, $V$, $R$, $i'$ and $z'$ optical observations with Subaru/Suprime-Cam,
reaching $B=28.4$, $R=27.4$ and $z'=26.2$ AB magnitude in 5 $\sigma$ (Wada et al. 2007 inpreparation)
and $J$, $K_S$ near-infrared observations with KPNO2.1 m/FMAMINGOS,
where the deepest regions reach to $J=21.85$ and $K_S=20.15$ Vega magnitude in 3 $\sigma$, respectively \citep{ima07}.
In this paper, since we focus on the $K_S$ -band selected high redshift galaxy populations, we adopt the catalog of
the region reaching to $K_S\sim20$ Vega magnitude in 3 $\sigma$.
The total effective area is 540 arcmin$^2$.

\section{Sample Selection of EROs, BzKs and DRGs}
\label{sec:Sample Selections of EROs, BzKs and DRGs}

\subsection{The ERO Sample}
\label{sub:The ERO Sample}
The selection of EROs depends on the depth of the photometry, the available filters and assumed color criteria.
In this paper, although EROs are defined as objects with $R-K_S>5.3$ following the classification scheme of \citet{poz00},
we also present an alternative ERO selection criteria of $R-K_S>5.0$ to compare our results with other published ERO counts.

\begin{figure}
\epsscale{.80}
\plotone{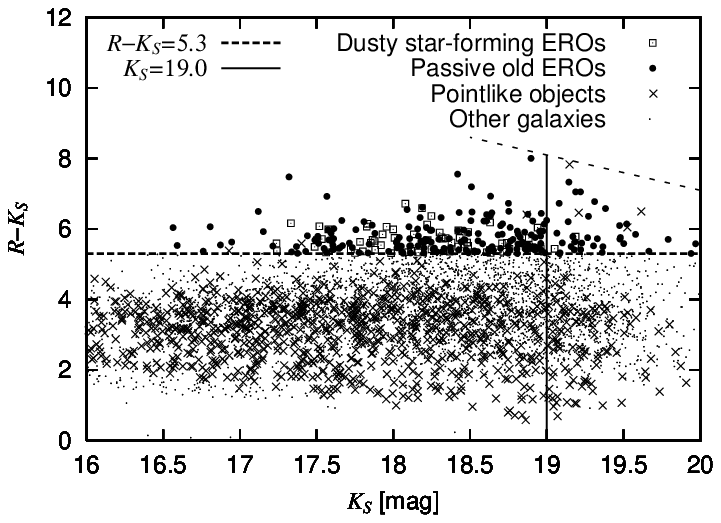}
\caption{$R-K_S$ color against $K_S$ magnitude in the $AKARI$ NEP field.
The open boxes and filled circles represent dusty star-forming and passive old EROs selected with the $R-K_S>5.3$ criterion (the horizontal dashed line).
The classification of dusty star-forming and passive old EROs is given in Figure \ref{figure:pEROvsdERO}.
The dots are other extended objects while the cross symbols are point-like objects.
The vertical solid line represents the magnitude limit at $K_S=19.0$.
The thin-dashed line shows the upper bound of the $R-K_S$ color due to the survey magnitude limit;
galaxies above this line are too faint to be detected in the $R$ -band ($R>27.0$ mag).}
\label{figure:ERO_select} 
\end{figure}

\begin{figure}
\epsscale{.80}
\plotone{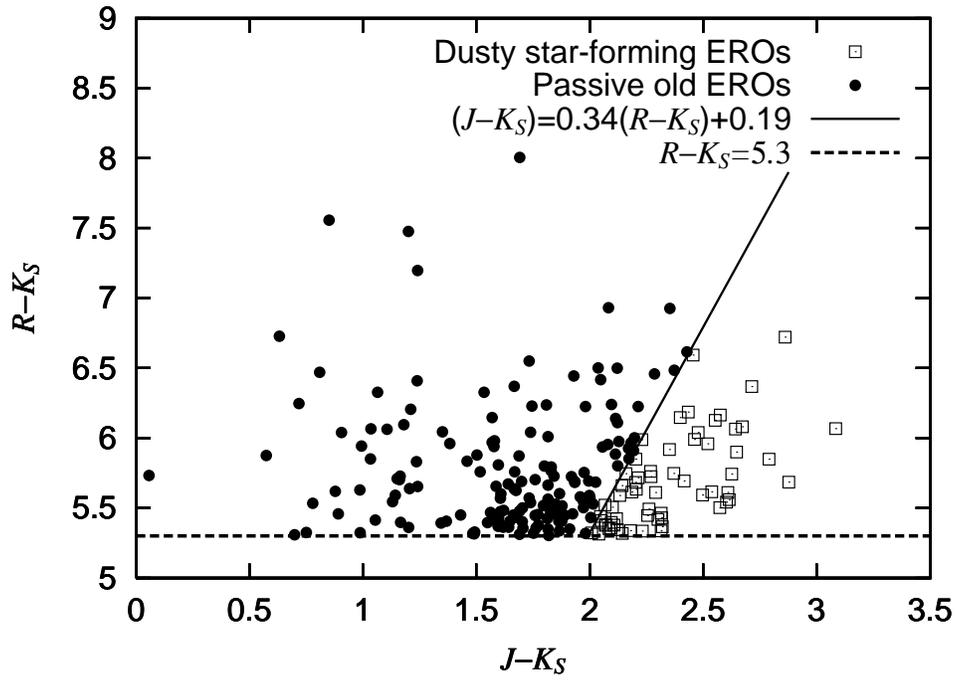}
\caption{$J-K_S$ vs. $R-K_S$ colors of EROs in the $AKARI$ NEP field.
The dusty star-forming ERO / passive old ERO separation is made based on the color discriminator from \citet{poz00}.
The open squares and filled circles represent dusty star-forming and passive old EROs, respectively.}
\label{figure:pEROvsdERO}
\end{figure}

Figure \ref{figure:ERO_select} shows the $R-K_S$ color versus $K_S$ magnitude for all detected objects in the $AKARI$ NEP field.
In total, we identify 308 (442) EROs with $R-K_S>5.3$ ($R-K_S>5$) to $K_S=19.0$.
We also calculated completeness correction factors as a function of $K_S$ magnitude by applying Monte Carlo
simulations to the real images (see \citet{ima07} for details of the method).
Above the magnitude of $K_S=19$ the catalogue appears more than 70\% complete.
Though the ERO candidates include the $R$ -band undetected sources, this is less than 4\% of the total ERO counts.
Some faint galaxies may be identified as point-like objects because their extended structure vanishes into the background noise.
The cross symbols in Figure \ref{figure:ERO_select} represent point-like objects
which are defined as less than 5.4 pixels (corresponding to 1.08'') in the $z'$ -band image
(a verification of the star and galaxy separation is described in \citet{ima07}).
Our result shows that two point-like objects in the magnitude range $17<K_S<19.0$ satisfy the ERO selection criteria of $R-K_S>5.3$.
Although, we do not include these objects in the ERO counts,
their contribution is $\sim0.6$\% of the total ERO counts and therefore will not significantly affect the results.
Figure \ref{figure:pEROvsdERO} shows the separation method of \citet{poz00}.
The extracted numbers of pERO and dERO are 167 and 141 to $K_S=19.0$, respectively. 

Note that an alternative selection method to segregate the pERO and dERO populations would be a direct measurement of their mid- \& far-infrared flux. Using $Spitzer$ \citet{ste06} observed the two archetypal extremely red objects  HR10 (dERO) and LBDS 53W091 (pERO) from the IRAC 3.6 $\mu$m band through to the MIPS 160 $\mu$m band. Although both sources were well detected in the four  shorter IRAC bands, only HR10 (i.e. a dusty star-forming galaxy) was detected in the MIPS 24 \& 70  $\mu$m bands. Obviously, mid- far-infrared detections of EROs provides a powerful alternative segregation criteria for dusty EROs and we will present similar $AKARI$ mid-infrared fluxes for our sample in a later work on completion of the $AKARI$ NEP survey.

\subsection{The BzK Sample}
\label{sub:The BzK Sample}
\begin{figure}
\epsscale{.80}
\plotone{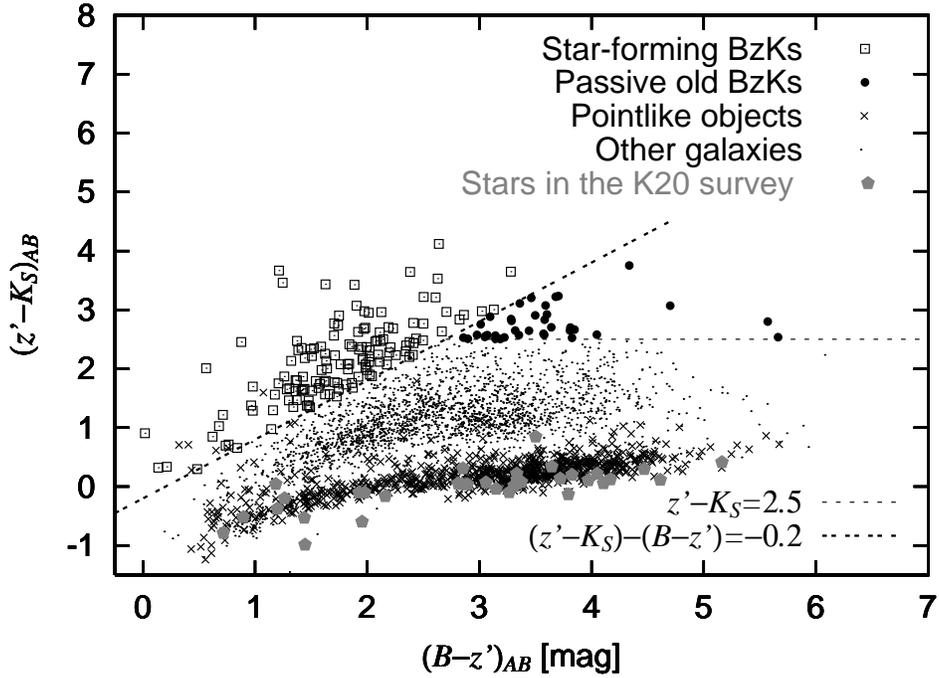}
\caption{$B-z'$ vs. $z'-K_S$ colors of $K_S$ -band selected objects in the $AKARI$ NEP field.
The discriminating color criterion is adopted from \citet{dad04}.
The open boxes and filled circles represent the star-forming and passive old BzKs, respectively.
The dots are other extended galaxies while the cross symbols are point-like objects.
The dots and two dotted lines represent the $(z'-K_S)_{AB}=2.5$ and $(z'-K_S)_{AB}-(B-z')_{AB}=-0.2$, respectively.}
\label{figure:BzK_select}
\end{figure}

We select sBzKs and pBzKs by using the color criteria of \citet{dad04}.
Figure \ref{figure:BzK_select} shows the $BzK$ color diagram of $K_S$ -band selected objects in the $AKARI$ NEP field.
In order to apply the $BzK$ selection criteria in a way consistent with \citet{dad04},
we normalize the stellar sequence in the $AKARI$ NEP survey to that of \citet{dad04}, resulting in a shift of
0.1 mag to fainter $K_S$ magnitudes.
From the results, we further selected 38 pBzKs and 137 sBzKs to $K_S=19.0$, above which the catalog is more than
70\% complete (see \citet{ima07} for details).
Seven  point-like objects also satisfy the sBzK criteria.
Since their contribution is less than 5\% of the total sBzK counts, we do not include them in the sBzK counts.

\subsection{The DRG Sample}
\label{sub:The DRG Sample}
\begin{figure}
\epsscale{.80}
\plotone{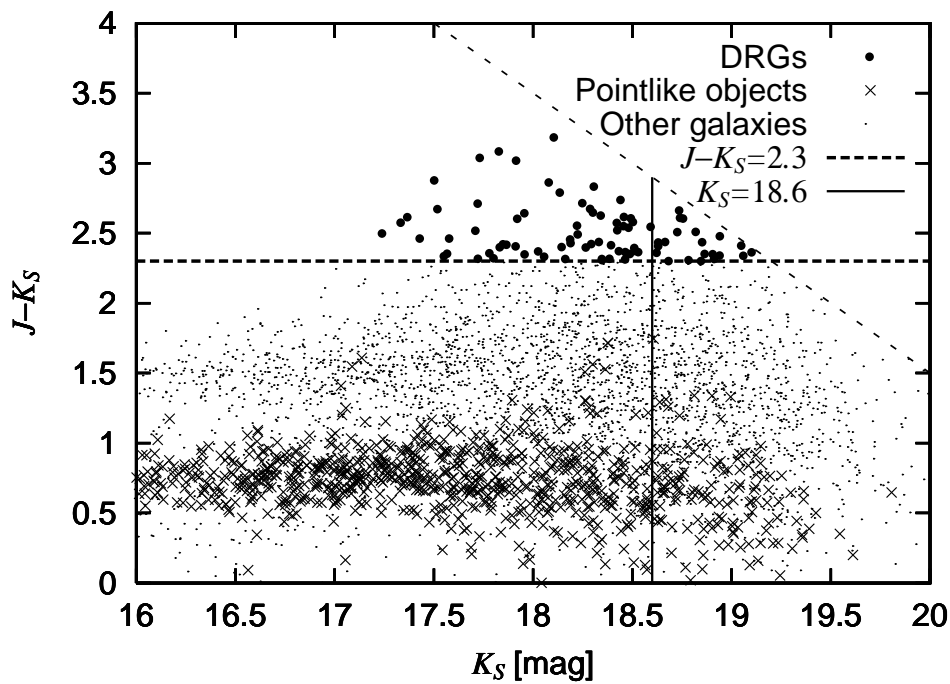}
\caption{$J-K_S$ colors against $K_S$ magnitude in the $AKARI$ NEP field.
Filled circles represent DRGs selected with the $J-K_S>2.3$ criterion while dots are other extended galaxies.
The cross symbols are point-like objects.
The dashed and solid lines represent the $J-K_S>2.3$ criterion and the magnitude limits at $K_S=18.6$, respectively.
The dotted line shows the upper limit of $J-K_S$ color due to the $J$ -band limit; galaxies above this line are too faint to be detected at $J$ -band ($J>21.5$ mag).}
\label{figure:DRG_select}
\end{figure}

We selected the DRG sample using the $J-K_S>2.3$ criterion \citep{fra03,van03}.
Figure \ref{figure:DRG_select} shows the $J-K_S$ color against $K_S$ magnitude for all detected objects in
the $AKARI$ NEP field.
The objects shown by the cross symbols are point-like objects.
For the DRG candidates, there is no contamination from point-like objects using the $J-K_S>2.3$ criterion.
However the $J$ -band images are not deep enough to find counterparts for all the $K_S$ -band selected objects.
We therefore restrict the DRG sample to be brighter than $K_S=18.6$. 
This threshold includes all DRGs bluer than $J-K_S\sim2.9$ which is almost the same criterion of \citet{fou06}
and provides 64 DRGs to $K_S=18.6$,
above which the catalog is 90\% complete (see \citet{ima07} for details).

\section{EROs, BzKs and DRGs counts}
\label{EROs_BzKs_and_DRGs_counts}
In this section, we present the $K_S$ -band number counts for the ERO, BzK and DRG populations (the relationship between th counts are discussed in \S \ref{sub:difference_of_counts}).
The raw numbers of these $K_S$ -band selected high redshift galaxy populations and completeness corrected counts
per square degree per magnitude are listed
in Table \ref{table:ERO_sample}, \ref{table:dpERO_sample}, \ref{table:dpBzK_sample} and \ref{table:DRG_sample}.

\begin{figure}
\epsscale{.80}
\plotone{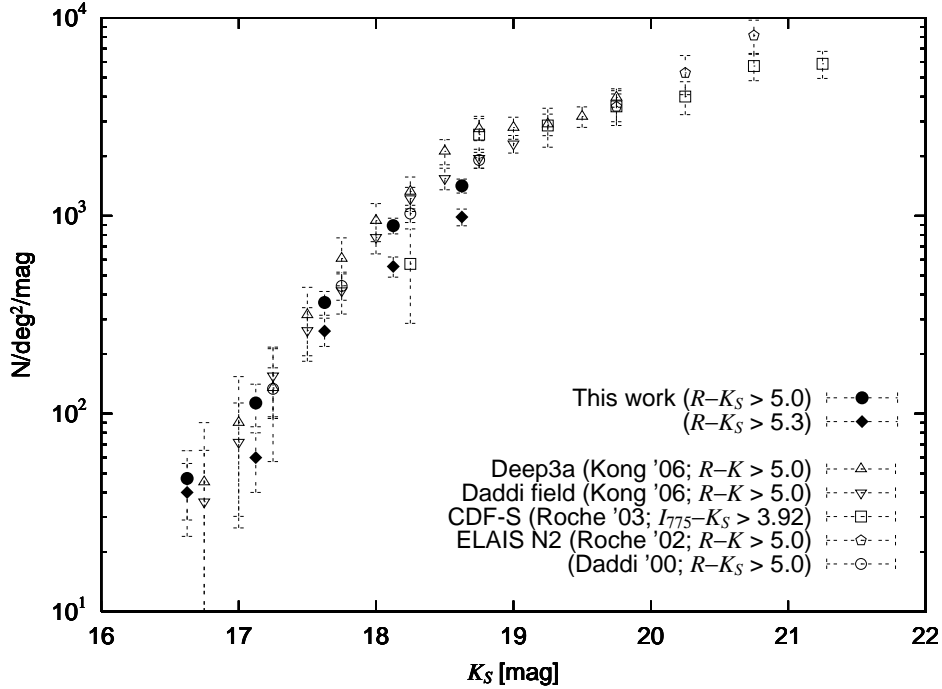}
\caption{Differential galaxy number counts for EROs.
The filled circles and diamonds are the counts in the $AKARI$ NEP field for EROs extracted according to the different selection criteria
$R-K_S>5.0$ and $R-K_S>5.3$, respectively.
Error bars are calculated from the number of galaxies in each magnitude bin using Poisson $\sqrt{N}$ counting statistics.
Other symbols in the figure correspond to the counts in the Deep3a and Daddi field \citep{kon06}, CDF-S \citep{roc03}, ELAIS N2 \citep{roc02}.
The open circles represent the results of \citet{dad00}.}
\label{figure:ERO_count}
\end{figure}

The ERO counts are presented in Figure \ref{figure:ERO_count}, which shows the results from
the different selection criteria along with a compilation of published ERO counts.
The ERO counts in the $AKARI$ NEP field are in good agreement with those of \citet{dad00}
where the EROs were selected using an identical criteria.
However our number counts in the faint magnitude range $K_S>18.5$ are slightly lower than those of \citet{roc03,kon06}
who utilized different filter bands to select the EROs.

\begin{figure}
\epsscale{.80}
\plotone{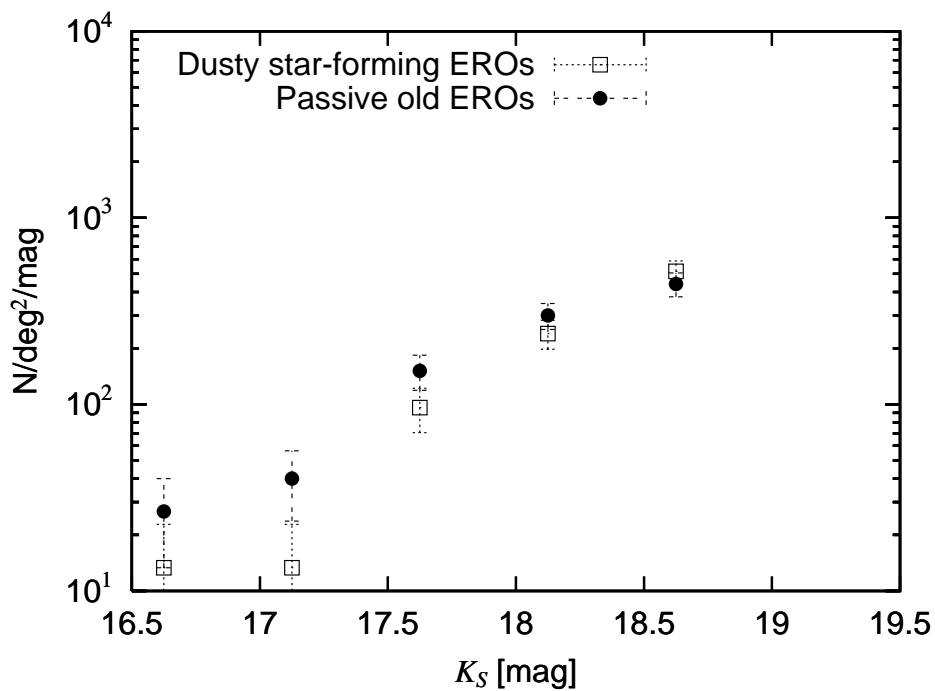}
\caption{Differential galaxy number counts for dusty star-forming and passive old EROs in the $AKARI$ NEP field based on the separation criteria given by \citet{poz00}.
The open squares and filled circles correspond to dusty star-forming and passive old ERO counts, respectively.
Error bars are calculated from the number of galaxies in each magnitude bin using Poisson $\sqrt{N}$ counting statistics.}
\label{figure:pdERO_count}
\end{figure}

Although there are many examples of total ERO counts and the segregation of the  two constituent populations in the literature \citep{roc02, roc03, miy03, vai04, kon06, sim06}, the individually segregated  source counts of both pERO \& dERO populations have yet to be coherently presented. In this work, we also present the observed counts of pEROs and dEROs separately in Figure \ref{figure:pdERO_count}.
In the bright magnitude range, the pERO counts are higher than that of the dEROs; by a factor of $\sim1.5$ at $K_S\sim17.6$.
This result is consistent with \citet{vai04} who concluded that the fraction of pEROs range from 65-80\% at $K<17$.

\begin{figure}
\epsscale{.80}
\plotone{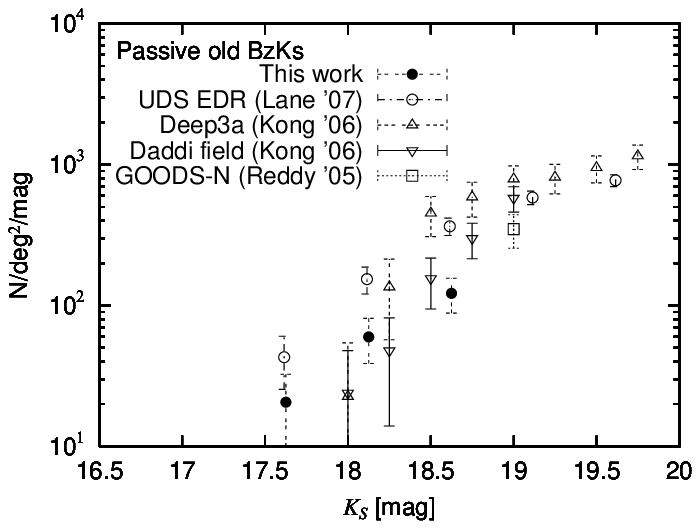}
\caption{Differential galaxy number counts for passive old BzKs.
The filled circles correspond to the counts in the $AKARI$ NEP field.
Error bars are calculated from the number of galaxies in each magnitude bin using Poisson $\sqrt{N}$ counting statistics.
The counts in the UDS EDR \citep{lan07}, Deep3a, Daddi field \citep{kon06} and GOODS-N \citep{red05} fields are also shown with other symbols.}
\label{figure:pBzK_count}
\end{figure}

\begin{figure}
\epsscale{.80}
\plotone{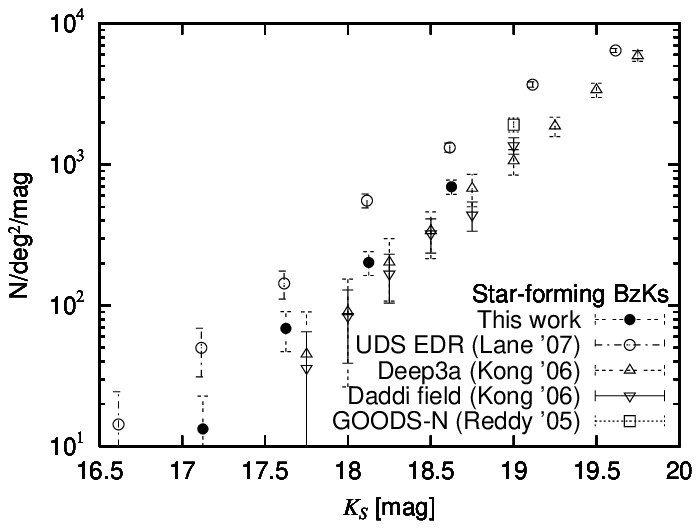}
\caption{Differential galaxy number counts for star-forming BzKs.
The filled circles correspond to the counts in the $AKARI$ NEP field.
Error bars are calculated from the number of galaxies in each magnitude bin using Poisson $\sqrt{N}$ counting statistics.
The counts in the UDS EDR \citep{lan07}, Deep3a, Daddi field \citep{kon06} and GOODS-N \citep{red05} fields are also shown with other symbols.}
\label{figure:sBzK_count}
\end{figure}

Figure \ref{figure:pBzK_count} and \ref{figure:sBzK_count} shows the number counts of pBzKs and sBzKs in
the $AKARI$ NEP field together with the results of \citet{red05,kon06}.
Our pBzK counts are broadly consistent with the counts in the Daddi field of \citet{kon06}
while the number counts in the Deep3a field of \citet{kon06} and UDS EDR field of \citep{lan07} are higher by
a factor of $\>3 $ at $K_S\sim18.5$. 

The discrepancies between the field to field counts could in some part be attributed to varying selection criteria, however another possibility is variation in the number densities due to the effects of cosmic variance. For small size fields and strongly 
clustered populations such as EROs, the uncertainty due to the cosmic variance can be as high as  30-40\% \citep{som04}, where as  larger survey areas such as the UDS EDR field of \citet{lan07} of 0.6~deg$^{2}$ are less susceptible with an estimated cosmic variance at the 3-4\% level over an interval $\delta z \sim$0.2 \citep{pea94}.

Combining the data with \citet{red05}, \citet{kon06} and \citet{lan07},
the slope of the pBzK counts shows a turn over at $K_S\sim19.0$.
On the other hand, our sBzK counts are in good agreement with \citet{kon06} in both the Deep3a and Daddi fields at
all magnitudes $17.5<K_S<18.8$ within the error bars.
The sBzK counts are more than 5.8 times higher than the pBzK counts at $K_S\sim18.6$ and show a steeper slope than
those of the pBzKs in the faint magnitude range.

\begin{figure}
\epsscale{.80}
\plotone{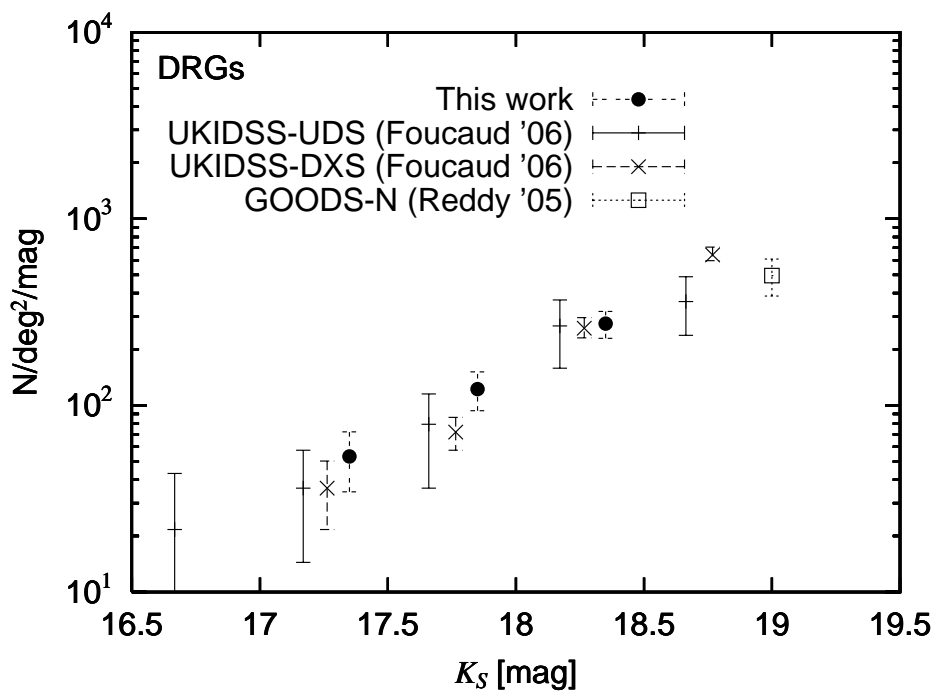}
\caption{Differential galaxy number counts for DRGs.
The filled circles are DRG counts in the $AKARI$ NEP field.
Error bars are calculated from the number of galaxies in each magnitude bin using Poisson $\sqrt{N}$ counting statistics.
Other symbols in the figure denotes the counts in UKIDSS-UDS, UKIDSS-DXS \citep{fou06} and GOODS-N \citep{red05} fields.}
\label{figure:DRG_count}
\end{figure}

Finally we present the DRG counts in Figure \ref{figure:DRG_count} together with the results of \citet{red05,fou06}.
The DRG counts are in good agreement with \citet{fou06} and the slope of the counts do not show a turnover to $K_S<18.5$.

\section{Discussion}
\label{sec:discussion}
We have presented ERO, BzK, DRG counts in \S \ref{EROs_BzKs_and_DRGs_counts}.
In this section, we discuss the relationship between, and the trend of the ERO, BzK and DRG count slopes (\ref{sub:difference_of_counts}), the results of comparing our counts with evolutionary models for the passive early-type galaxy population (\ref{sub:comparing_with_model}) and the dERO star-forming population (\ref{sub:comparing_with_dusty_model}). 

\subsection{Difference between the ERO, BzK and DRG counts}
\label{sub:difference_of_counts}
As previously indicated by \citet{roc03} in other ERO fields, the ERO counts in the $AKARI$ NEP field (Figure \ref{figure:ERO_count}) 
also clearly show a slope change at $K_S\sim18.0$ below which the counts exhibit a shallower slope.
On the other hand, the DRG counts show no sign of any flattening of the slope at $K_S<18.5$,  although it should be noted that the DRG counts in the $AKARI$ NEP field are drawn from limited data across only three magnitude bins.

However a shallower slope has been reported at fainter magnitudes from the deeper DRG counts \citep{fou06,gra061,kaj07} and
\citet{fou06} suggest a break feature in the slope at $K_{AB}\sim20.5$.
Although EROs and DRGs both include dusty star-forming and passive old populations,
these results suggest that the red galaxies contribute less significantly to the galaxy counts in the fainter magnitude range.

The pERO counts show a turn over at $K_S\sim18.5$
whilst the pBzK counts show a turn over at $K_S\sim19.0$, shifting to slightly fainter magnitudes.
This is reasonable since the pBzK criterion is generally sampling a higher redshift domain than the pERO criterion.
Since the redshift distributions of the pERO and pBzK counts are well-defined,
these populations are independently selected passive old galaxies in different redshift slices.
Therefore the pERO and pBzK criteria can provide a good opportunity to investigate the evolution of early-type galaxies.
We will investigate the difference between these counts of passive old samples and produce a number count model for
the early-type galaxies in next subsection.

\subsection{Comparison with early-type galaxy count models}
\label{sub:comparing_with_model}
In this section, we construct a number count model for the early-type galaxies in order to explain the pERO and pBzK counts.
We primarily adopt the number count model based on the formulation of \citet{gar98}.
This model incorporates the evolution of the LF for each galaxy population within a backward evolution framework. 
Therefore, we may control the evolution of each galaxy type using the type-dependent LF.

The number count model depends mainly on the three parameters of the \citet{sch76} LF,
\begin{equation}
\phi(L)dL=\phi^*\left(\frac{L}{L^*}\right)^\alpha\exp\left(-\frac{L}{L^*}\right)\frac{dL}{L^*},
\end{equation}
with
\begin{equation}
\frac{L}{L^*}=10^{0.4(M^*-M)}.
\end{equation}
The parameter $M^*$ is the characteristic magnitude and the parameter $\alpha$ is the faint-end slope of the LF.
Although the parameter $\phi^*$ is a normalization coefficient with dimensions of the number density of galaxies,
we re-define $\phi^*$ as follows, in order to include a parameterization of evolution in the number density,
\begin{equation}
\label{equation:density_evolution_parameter}
\phi^*(z)=\phi^*(0)(1+z)^m.
\end{equation}
where $m$ is a free parameter controlling the number density evolution. In the case of $m<0$, the number density is decreasing to higher redshift (hereafter `negative density evolution').

In the past decade, there have been a number of determinations of the NIR local LF of galaxies
based on both photometric \citep{bal01,kas03,cap05,sar06} and spectroscopic redshifts \citep{gla95,hua03,poz03}.
From optical surveys, it has also been shown  that each galaxy type has a characteristic LF shape  \citep{fol99,mad02,nak03,dah05}.
However LFs determined with optical surveys may be affected by star-formation activity and dust extinction,

we refer to the type-dependent $K$-band luminosity function (LF) of \citet{koc01}
based on the 2MASS Second Incremental Release Catalog of Extended Sources \citep{jar00}.
However \citet{col01} point out that the $K$-band photometry of \citet{lov00} has better signal-to-noise and resolution than the 2MASS images and enables more accurate 2MASS magnitudes to be measured than the original default magnitudes \citet[see also]{fri05}.
We therefore normalized the characteristic magnitude $M^*_K$ of the number count model at $K<13.25$ of the 2MASS counts measured by \citet{fri05} using a similar magnitude estimator to that of \citet{col01}. The three parameters for the LF of early-type galaxies are shown in Table \ref{table:early_type_parameter}.

We use the spectrophotometric population synthesis models of \citet{bru03} with a Salpeter stellar initial mass function
to compute the evolution of the galaxy luminosity as a function of the age, which we refer to as our baseline Pure Luminosity Evolution (PLE) case.
This theoretical spectral energy distribution (SED) model is ideal for the approach to number count modelling
within the backward evolution framework
because it returns the SED of a galaxy with any star formation history as a function of cosmic time for a given formation epoch.
The basic characteristics of the SED of early-type galaxies i.e., metalicity and star formation rate, assuming pure luminosity evolution, are also listed
in Table \ref{table:early_type_parameter}.

The galaxy formation epoch is an important model parameter, as
any delay in the galaxy formation epoch will shift the count slope towards brighter magnitudes and lower number densities.
It is, however, poorly understood and rather loosely constrained.
\citet{hol04} reported a lower limit for the elliptical galaxy formation epoch from a study of the color-magnitude relation
of elliptical and S0 galaxies in six clusters obtained using imaging  on the HST.
They reported the oldest stars in the elliptical galaxies appear to have formed at $z>3$. 
We therefore calculate our number count models for the early-type galaxy population for three different formation epochs at $z_{form}=4$, 5 and 6.

\begin{figure}
\epsscale{.80}
\plotone{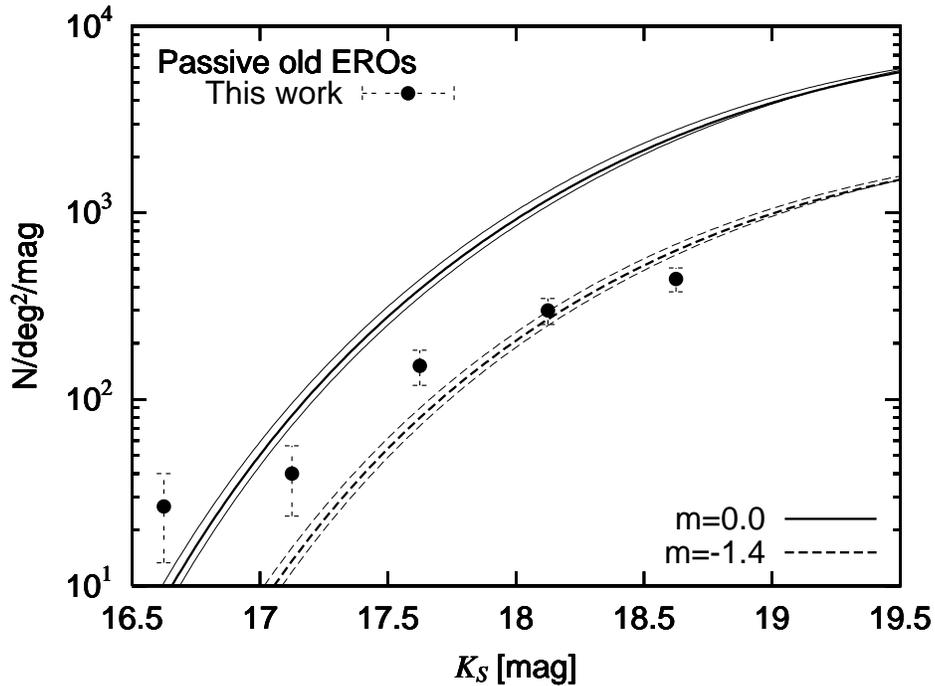}\\
\caption{$K_S$ -band number count predictions for the passive old EROs.
Passive old EROs in the $AKARI$ NEP field are shown as filled circles.
The bold solid and dashed lines represent the PLE and the PLE plus density evolution model for a formation epoch $z_{form}=5.0$.
The accompanying upper and lower thin lines correspond to the different formation epochs of $z=3,4,5$, respectively.}
\label{figure:pERO_model}
\end{figure}

We first compare the calculated early-type number counts with the observed pERO counts as shown in Fgure \ref{figure:pERO_model}.
The effect of the strong negative number density evolution is seen as a shift of the counts to lower number densities with magnitude.
In the bright magnitude range $K_S<17.0$, since the pEROs sample is small in number,
the simple PLE model without density evolution (corresponding to $m=0$) cannot be ruled out.
However in the fainter magnitude range $K_S>17.0$, the simple PLE model is clearly inconsistent with the pERO counts:
the faintest pERO counts at $K_S\sim19.1$ are more than 6.7 times lower than the PLE models with $z_{form}=5$.
Note the pERO counts are higher than the negative density evolution model with $m=-1.4$,
this is because the brighter samples of pEROs contain many low redshift galaxies \citep[e.g.] {vai04}.
We conclude that the pERO counts favor the negative density evolution models.

We also compare the early-type number count model with the observed pBzK counts.
The simple PLE models with $m=0$ exceed the observed counts at all magnitudes
while the negative density evolution models with $m=-1.4$ agree well with the bright pBzK counts in the $AKARI$ NEP field
and the deeper pBzK counts of \citet{red05,kon06}.

The number count model with negative number density evolution ($m=-1.4$) shows better fits for both the pERO and pBzK counts.
Since independently selected passive old samples both show signs of negative density evolution,
these results indicate that the overall early-type galaxy population may exhibit negative density evolution.

\begin{figure}
\epsscale{.80}
\plotone{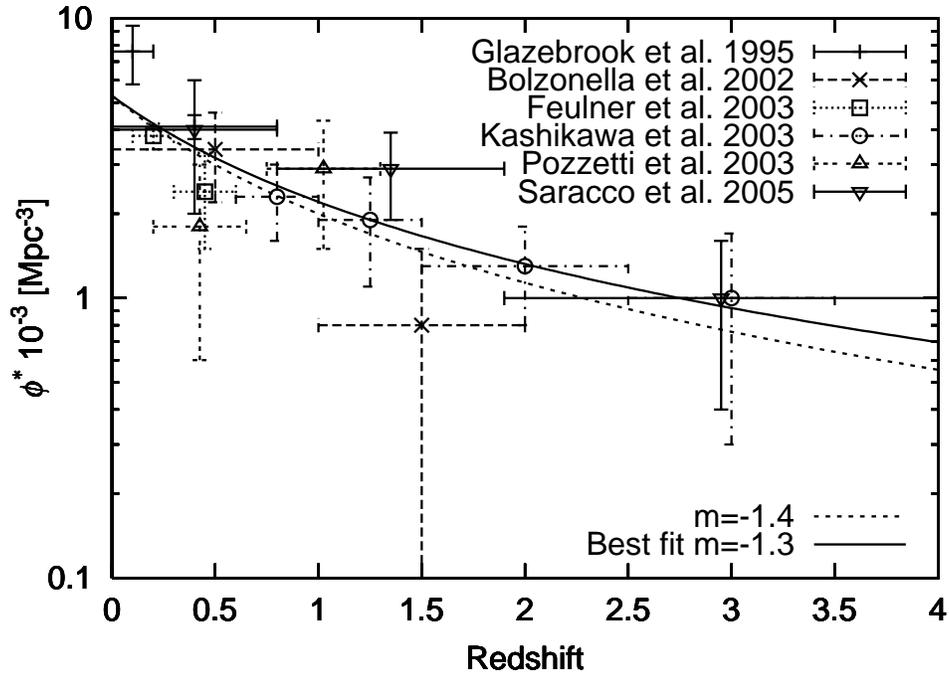}\\
\caption{Evolution of $K_S$- band luminosity function.
The dotted and dashed lines represent the result of the negative density evolution given by eq. \ref{equation:density_evolution_parameter}:
$m\sim-1.3$ and -1.4, respectively}
\label{figure:LF_evolution}
\end{figure}

Indeed negative density evolution has also been reported from studies of the evolution of the $K$ -band LF at various redshifts: $z<1$
\citep{gla95,feu03}, $z<2$ \citep{bol02,poz03}, $z<3.5$ \citep{kas03} and $z<4$ \citep{sar06}.
We compile these results in Figure \ref{figure:LF_evolution}, and compare with the negative density evolution model.
The best fit is obtained for $m\sim-1.3$.
Note that \citet{poz03} report a contrary behavior, however, their large errors due to the limited sample size do not rule out a decrease in the comoving number density with redshift.
To summarize, our counts are consistent with strong negative density evolution  ($m=-1.4$) in the early-type galaxy population color selected (pERO and pBzK) sample, whiles the total rest-frame $K$ -band LF  also exhibits negative density evolution with redshift with a best fit parameter $m\sim-1.3$.

%%  *********************************************************************************************
%%  *****************  Chris's Dusty ERO contribution - START * *********************
%%  *********************************************************************************************

\subsection{Comparison with dusty star-forming galaxy count models}
\label{sub:comparing_with_dusty_model}

In this section, we will construct a simple number count model for dusty  star-forming galaxiesto represent the dERO counts. 
%% Note that the star-forming BzK galaxies are intrinsically blue sources and may comprise a none dusty population {\bf NEED~REFERENCES~AND~TO~CHECK~THIS}. Therefore we do not attempt to model the   star-forming BzK galaxies in our current work.
We assume that the nature of the dERO's are high redshift starbursts with their redness being due to the extreme dust extinction within the host galaxies. It follows that if indeed the dEROs are significant sites of star-formation then these sources should also exhibit significant emission at infrared-submillimetre wavelengths \citep{elb02}. 

In recent years a great deal of effort has gone into estimations of the fraction of dEROs making up the far-infrared--submillimetre counts, however these estimates have been hampered by differing selection criteria. Estimates of the fraction of dEROs in the bright ($S_{850\mu m} > 2$mJy) submillimeter population have been made in the SCUBA-CUDSS fields where 10-20$\%$ of sources were identified as EROs \citep{web03},
although \citet{cle04} found a higher fraction of 8/14 in the CUDSS 3hr field
sources could be classified as EROs. In the SCUBA-8mJy fields, 6/30 sources (6/17 with optical identifications) were identified as dEROs  \citep{ivi02}.
More recently, \citet{tak07} have estimated a dERO contribution of 10$\pm$7$\%$ (for $K_S<20.3$) in the SCUBA-SHADES survey \citep{mor05} although this fraction may increase to fainter $K$ magnitudes.
However within the bright submillimetre fields, the average source density of near-infrared selected (p+d) EROs is $\sim$18,000 per square degree, i.e.
significantly higher than the source density of bright submillimtre galaxies ($\sim$thousands per square degree).
Therefore, for the dEROs to be explained by the submillimeter source population there must exist a larger contribution at fainter submillimetre fluxes, and indeed,
deeper submillimetre observations ($S_{850\mu m} < $2mJy) have implied a much higher fraction of EROs. Using  the HDF-N SCUBA super-map,
\citet{pop05} discovered a much higher contribution of EROs than in the brighter submillimetre surveys and moreover classified all submillimetre detected EROs as dEROs.
At submillimetre fluxes of $S_{850\mu m}\sim$1mJy,
the source density of submillimetre sources becomes sufficiently high to account for all EROs.
To investigate the contribution of EROs to the fainter submillimetre population, \citet{knu05} performed submillimetre stacking analysis of red galaxies around distant clusters finding an average flux of $S_{850\mu m} \sim $1mJy  and an estimated contribution to the total flux from faint sources between 0.5 $ < S_{850\mu m} <$ 5mJy of 50$\%$. Similar studies have yielded consistent results with  average fluxes of 1.58$\pm$0.13mJy and contributions to the emission over the 
0.5 $ < S_{850\mu m} <$ 5mJy flux range  of 50$\%$ \citep{weh02}, and 0.4$\pm$0.07mJy, with a contribution to the total emission over the 0.5 $ < S_{850\mu m} <$ 5mJy range of 10$\%$ \citep{web04} respectively. The differences are most likely due to the slightly different ERO depth and selection criteria and field to field variations due to the strong clustering nature of dEROs.

Under the above assumptions, we model the $K$-band dusty star-forming ERO population as a fraction of the dusty far-infrared -- submillimetre population.  We assume the submillimetre luminosity function of  \citet{ser05} that was derived from a combination of the IRAS point source catalogue \citep{sau00} and the SCUBA Local Galaxy Survey \citep{dun00}. 
The luminosity function is a double power law (somewhat broader than the traditional Schecter function in the higher luminosity regime) and is defined as;

\begin{equation}
\phi(L)dL=\frac {\phi^*}{(\frac{L}{L^*})^{\beta}+(\frac{L}{L^*})^{\gamma}}dL
\end{equation}

where $L^*$ is the characteristic luminosity (with corresponding characteristic magnitude,  $M^*$, given by $L/L^*$=10$^{0.4(M^*-M)}$ and $\phi^*$ the number density normalization coefficient at $\phi(L^*)$. The parameters, $\beta$ \&  $\gamma$ control the faint and bright end slopes of the luminosity function. The adopted luminosity function parameters are shown in Table~\ref{table:starforming_type_parameter}. We define three bolometric luminosity regimes for our luminosity function of $L/L_{\odot}>$10$^{10}$, 10$^{11}$, 10$^{12}$ corresponding to star-forming infrared galaxies (STFG), luminous infrared galaxies (LIRG) \& ultra-luminous infrared galaxies (ULIRGs) respectively. These populations are modelled within the evolutionary framework of \citet{pea08} which includes both luminosity evolution ($k$) and density evolution ($g$) as double power laws, of the form $(1+z)^{k,g}$, rising steeply to a redshift of approximately one and shallower thereafter. The evolutionary strength and form is defined by a pair variables for both the luminosity and the density evolution: the power index from redshift zero to redshift  unity, $k_{1}$,$g_{1}$ and an additional power index, $k_{2}$,$g_{2}$, to a higher redshift z$_{2}\sim$3.0, after which the evolutionary power is assumed to decrease exponentially (see \citet{pea08} for explicit details of the model). The parameterization of the power laws are given in Table \ref{table:dEROevolution}.

Our source SEDs are selected from the spectral libraries of \citet{tak03a,tak03b} which provide excellent fits to the SEDs of submillimetre galaxies from submillimetre to near-infared wavelengths \citep{tak04}. These models assume a Salpeter IMF of index 1.35 from 0.1-60M$_{\sun}$  and parameters for the star formation timescale (0.1Gyr), metallicity (assumed to be 0.1), the compactness of the starburst region ($\Theta$) and the starburst age. The variation in the starburst SED is explained by the difference in the starburst age and the compactness of the starburst region. The starburst age ranges from 0.01-0.6Gyr. The compactness factor $\Theta$ ranges from 0.3 -- 5.0 and can be considered as a measure of the optical depth. In order to select SEDs appropriate for our modelling of the dERO population we identify those SEDs from within the spectral library that specifically have red colours  $R-K >$5.3 in the redshift range 1-2. The selected SEDs for our three populations are tabulated in Table~\ref{table:starforming_type_parameter}. Note that for our $L/L_{\odot}>$10$^{12}$ population (i.e. ULIRGs) we adopt the model fit to the archetypical dERO HR10 (ERO J164502+4626.4, \citet{hu94}) which has been shown to be a distant Arp220 type ULIRG analogue \citep{elb02}. 

\begin{figure}
\epsscale{.80}
\plotone{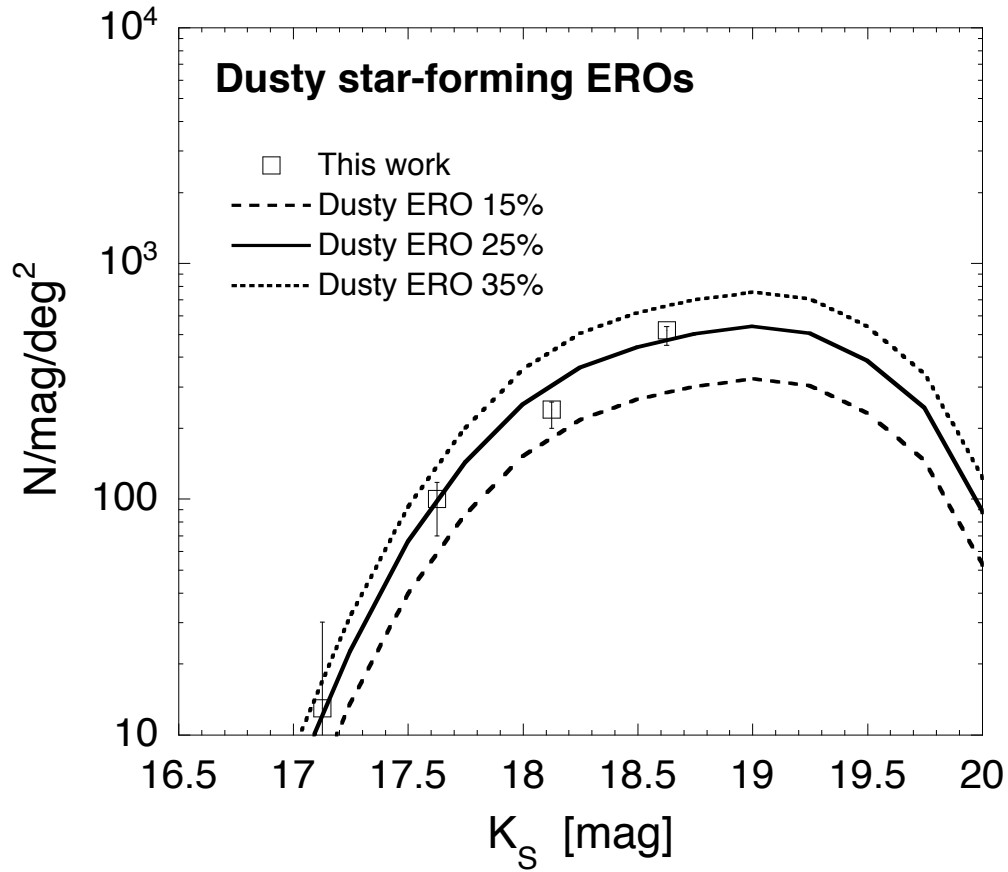}\\
\caption{$K_S$ -band number count predictions for the dusty star-forming EROs.
The lines represent different assumed sub-millimetre population fractions.
The bold solid corresponds to a contribution of 25 percent of dEROs to the submillimetre population and the upper dotted and lower dashed lines represent contributions of 35 percent and 15 percent respectively.
Dusty star-forming EROs in the $AKARI$ NEP field are shown as open squares.}
\label{figure:dERO_model}
\end{figure}

\begin{figure}
\epsscale{.80}
\plotone{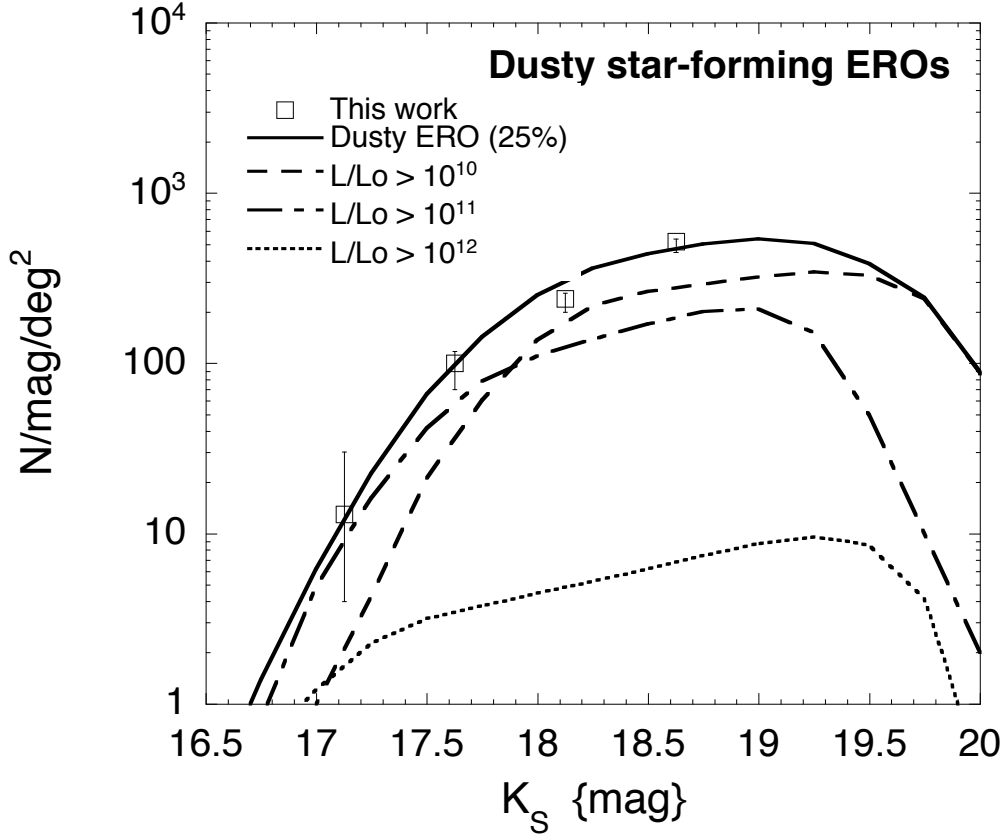}\\
\caption{$K_S$ -band number count predictions for the dusty star-forming EROs.
The lines represent different assumed galaxy populations corresponding to the bolometric luminosity regimes of star-forming galaxies (STFG) $L/L_{\odot}>$10$^{10}$ (dashed), luminous infrared galaxies (LIRG) $L/L_{\odot}>$10$^{11}$ (dot-dashed),  \& ultra-luminous infrared galaxies (ULIRGs) $L/L_{\odot}>$10$^{12}$ (dotted).
The bold solid corresponds to the total contribution of dEROs  (the same as in Figure \ref{figure:dERO_model}).
Dusty star-forming EROs in the $AKARI$ NEP field are shown as open squares.}
\label{figure:dERO_parts}
\end{figure}

Using the above luminosity function and SEDs we model the observed counts of dEROs in the $AKARI$ NEP field by assuming ERO fractions of the submillimetre population of 15, 25 \& 35 percent respectively. The resulting counts are shown in Figure \ref{figure:dERO_model}. The $K_S$-band dERO number counts are well fitted by a submillimetre evolutionary model assuming that around 25 percent of the sub-millimetre galaxy population to faint fluxes are dEROs. The 35 percent model over-predicts the dERO counts at $K_S$ magnitudes fainter than 17, whilst the 15 percent model cannot account for the fainter dERO source counts at  $K_S>$18.5.
In Figure \ref{figure:dERO_parts} we also plot the individual components (STFG L/L$_{\odot}>$10$^{10}$, LIRG L/L$_{\odot}>$10$^{11}$, \& ULIRG  L/L$_{\odot}>$10$^{12}$, see Table \ref{table:starforming_type_parameter}) corresponding to our best fit  25 percent model given in Figure  \ref{figure:dERO_model}. In our best fit 25 percent model, the bulk of the brighter ($K_S<$19) counts originate from the  $L/L_{\odot}<$10$^{12}$ population, i.e. that HR10 type dEROs do not contribute significantly to the dERO source counts. Note that this result is consistent with the sub-mm stacking of ULIRGs of \citet{moh02} and the work of \citet{sma99} who concluded that the contribution of HR10-like dEROs to the submillimetre population was constrained to 10-20 percent. The brighter counts at ($K_S<$18) are composed of LIRG sources with luminosities 10$^{11}< L/L_{\odot}<$10$^{12}$ along with a non-negliable fraction of weaker L/$<$10$^{11}L_{\odot}$  star-forming galaxies. We also predict that by $K_S>$19.5, the dERO source counts are dominated by the faint submillimetre population of moderate star-forming galaxies. Note that if this is the case, then the fainter dEROs are examples of less luminous systems responsible for a large fraction of the star-formation rate at high redshift. We also predict that, corresponding to the decline of the brighter systems,  the counts of dEROs should turn over in the magnitude range $K_S$=19-20. 

\section{Summary}
\label{sec:summary}
In order to seek clues to understand the origin of the change in the $K$ -band count slope and the evolution of high redshift galaxy populations, we have investigated the number counts of $K$ -band selected high redshift galaxy populations: EROs, BzKs and DRGs. Moreover we have further segregated these populations into their passive elliptical and dusty evolving star-forming constiuents.
Our catalogue in the $AKARI$ NEP field contains 167 pEROs and 141 dEROs ($K_S<19.0$), 38 pBzKs and 137 sBzKs ($K_S<19.0$), 64 DRGs ($K_S<18.6$).
From these large samples, we confirmed that the number counts of both the pEROs and pBzK have shallower slopes at fainter magnitudes and turnovers at $K_S\sim18.5$ and $K_S\sim19.0$, respectively.
On the other hand, the sBzK counts show a steeper slope which dominates in the fainter magnitude range as shown by \citet{kon06, hay07}.
These results strongly support that the passively evolving population contributes less to the galaxy counts at fainter magnitudes.

For the passively evolving pERO and pBzKs populations, from a comparison with a backward evolution model allowing number density evolution, we conclude that the PLE model without density evolution is at odds to explain the observed counts at fainter magnitudes,
whilst the model including negative density evolution is consistent with the observed counts of the passively evolving populations.

We have also modelled the dERO population with a simple backward evolutionary model assuming that the dERO population
can be explained by the source densities of the far-infrared --
submillimetre populations. Our best ft to the dERO source
counts is found by assuming a 25 percent contribution from submillimetre
star-forming galaxies with the majority of brighter $K_S$ -band detected
dEROs having luminous, as opposed to  ultra-luminous counterparts. With such a significant contribution by dEROs to the fraction of dusty galaxies at far-infrared--submillimetre wavelengths, we may expect a correspondingly significant contribution, from this population, to the cosmic star-fromation history of the Universe.

Further investigation of these high redshift colour selected populations (ERO, BzK, DRG) will not only produce constraints on the formation epoch and evolution of giant galaxy systems but also provide a bridge to connect the optical and infrared star-formation history of the Universe. A discussion of the $AKARI$ mid-infrared emission from our sample will be given in a forthcoming work.

%The effect of the negative density evolution shifts the turnover of the count slope to lower number densities and the bright magnitudes ($K_S\sim18$)
%where the observed flattering in the $K_S$-band counts begins.

\begin{figure}
\epsscale{.80}
\plotone{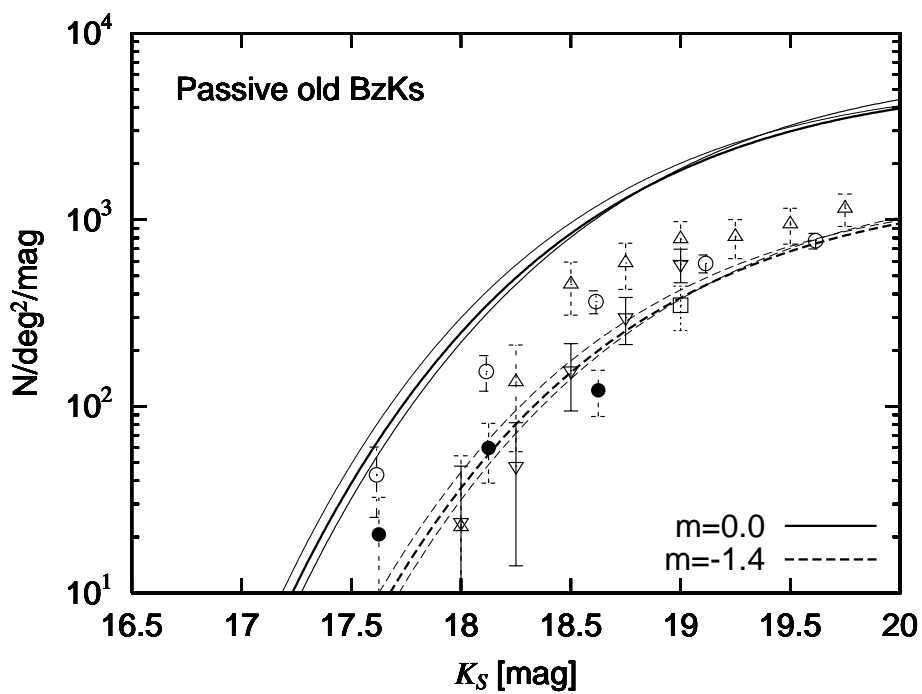}\\
\caption{$K_S$ -band number count predictions for the passive old BzKs.
Symbols for observed counts are the same as those in Figure \ref{figure:pBzK_count}.
The bold solid and dashed lines represent the PLE and the PLE plus density evolution model for a formation epoch $z_{form}=5.0$.
The accompanying upper and lower thin lines correspond to the different formation epochs of $z=3,4,5$, respectively.}
\label{figure:pBzK_model}
\end{figure}

%% In this first example, note that the \tabletypesize{}
%% command has been used to reduce the font size of the table.
%% We also use the \rotate command to rotate the table to
%% landscape orientation since it is very wide even at the
%% reduced font size.
%%
%% Note also that the \label command needs to be placed
%% inside the \tablecaption.

%% This table also includes a table comment indicating that the full
%% version will be available in machine-readable format in the electronic
%% edition.

\clearpage
\begin{table}
\begin{center}
\caption{The number of ERO samples in the $AKARI$ NEP field \label{table:ERO_sample}}
\begin{tabular}{ c c c c c c c c c c }
\tableline\tableline
 $K_S$ bin center	& \multicolumn{3}{c}{$R-K_S>5$}	& \multicolumn{3}{c}{$R-K_S>5.3$} \\
			& N$_{raw}$	& N$_{corr}$	& Error	& N$_{raw}$	& N$_{corr}$	& Error \\
 (1)		& (2)				& (3)				& (4)		& (5)				& (6)				& (7)	\\ \tableline
 16.625	& 47				& 47				& 18		& 40				& 40				& 16	\\
 17.125	& 110			& 110			& 30		& 53				& 53				& 19	\\
 17.625	& 350			& 36				& 50		& 240			& 250			& 40	\\
 18.125	& 790			& 890			& 80		& 490			& 550			& 60	\\
 18.625	& 1000			& 1420			& 120	& 700			& 990			& 100 \\
\tableline
\end{tabular}
\tablecomments
{Effective area of NE and SE fields is 540 arcmin$^2$.
Unit of the counts is per square degree per magnitude.\\
(2) and (5) Raw ERO counts.\\
(3) and (6) Completeness corrected counts.\\
(4) and (7) Poisson $\sqrt{N}$ statistics.}
\end{center}
\end{table}

\begin{table}
\begin{center}
\caption{The number of sample of dusty star-forming and passive old EROs \label{table:dpERO_sample}}
\begin{tabular}{ c c c c c c c }
\tableline\tableline
 $K_S$ bin center &	\multicolumn{3}{c}{Passive old EROs}	& \multicolumn{3}{c}{Dusty star-forming EROs} \\
						& N$_{raw}$	&N$_{corr}$	& Error	& N$_{raw}$	&N$_{corr}$	& Error	\\
  (1)					& (2)				& (3)		   		& (4)		& (5)				& (6)				& (7)		\\ \tableline
 16.625				& 27				& 27				& 13		& 13				& 13				& 9		\\
 17.125				& 40				& 40				& 16		& 13				& 13				& 9		\\
 17.625				& 150			& 150			& 30		& 90				& 100			& 30		\\
 18.125				& 270			& 300			& 50		& 210			& 240			& 40		\\
 18.625				& 310			& 440			& 60		& 370			& 520			& 70		\\
\tableline
\end{tabular}
\tablecomments
{Description of the columns are the same as in the footnote of Table \ref{table:ERO_sample}.}
\end{center}
\end{table}

\begin{table}
\begin{center}
\caption{The number of sample of star-forming and passive old BzKs \label{table:dpBzK_sample}}
\begin{tabular}{ c c c c c c c }
\tableline\tableline
 $K_S$ bin center & \multicolumn{3}{c}{Passive old BzKs}	& \multicolumn{3}{c}{Star-forming BzKs} \\
						& N$_{raw}$	&N$_{corr}$	& Error	& N$_{raw}$	&N$_{corr}$	& Error	\\
  (1)       			& (2)				& (3)  	 		& (4)		& (5)				& (6)				& (7)		\\ \tableline
 17.125				& ---			& ---			& ---	& 13				& 13				& 9		\\
 17.625				& 20				& 21				& 12  	& 70				& 70				& 20		\\
 18.125				& 50				& 60				& 20 		& 180			& 200			& 40		\\
 18.625				& 90				& 120			& 30		& 490			& 690			& 80		\\
\tableline
\end{tabular}
\tablecomments
{Description of the columns are the same as in the footnote of Table \ref{table:ERO_sample}.
Numbers in parenthesis include the point-like objects that also satisfy the BzK color selection criteria of \citet{dad04}}
\end{center}
\end{table}

\begin{table}
\begin{center}
\caption{The number of DRGs sample \label{table:DRG_sample}}
\begin{tabular}{ c c c c }
\tableline \tableline
 $K_S$ bin center & \multicolumn{3}{c}{$J-K_S>2.3$} \\
				& N$_{raw}$	&N$_{corr}$	& Error	\\
  (1)   	    & (2)				& (3)   			& (4)	\\ \tableline
 17.35		& 53				& 53				& 19	\\
 17.85		& 120			& 120			& 30	\\
 18.35		& 250			& 270			& 50	\\
\tableline
\end{tabular}
\tablecomments
{Description of the columns are the same as in the footnote of Table \ref{table:ERO_sample}.}
\end{center}
\end{table}

\begin{table}
\begin{center}
\caption{Adopted SED model and local LF parameters in the number count for early-type galaxies models \label{table:early_type_parameter}}
\begin{tabular}{ c c c c c }
\tableline \tableline
 Metalicity			& Star Fomation Rate	& M$_{K}^*$	& $\phi^*$	& $\alpha$ 			\\
 Z/Z$_{\odot}$	& Gyr						& 					& Mpc$^{-3}\times10^{-3}$ &	\\
  (1)       			& (2)							& (3)   			& (4)			& (5) 						\\ \tableline
 2.5					& exponential $\tau =0.3$		& -24.0			& 1.3			& -0.85					\\
\tableline
\end{tabular}
\tablecomments
{(2) Star formation and the $e$-folding time assuming Pure Luminosity Evolution (PLE).}
\end{center}
\end{table}

\begin{table}
\begin{center}
\caption{Adopted  local LF and SED parameters for star-forming galaxy number count models 
\label{table:starforming_type_parameter}}
\begin{tabular}{ c c c c c }
\tableline \tableline
\multicolumn{5}{c}{Model SED parameters}            \\      
Bolometric Luminosity	& starburst age	& IMF	&  $\Theta$	&  Extinction Curve   \\
L/L$_{\odot}$					& Gyr	& 			&			&      \\
  (1) 						& (2)		& (3)		& (4)		& (5)  \\ 
  $>$10$^{10}$					& 0.6		& 1.35	& 1.0		& SMC  \\ 
  $>$10$^{11}$             			& 0.6		& 1.35	& 0.9		& SMC  \\ 
  $>$10$^{12}$           			& 0.5		& 1.35	& 0.3		& SMC  \\ 
  \tableline
\multicolumn{5}{c}{Model luminosity function parameters}    \\              
        & L$_{submm}^*$      &  $\phi^*$ &           $\beta$                &  $\gamma$ \\
	&  lg(L/L$_{\odot}$) &  Mpc$^{-3}\times10^{-3}$&  &                \\
Serjeant \& Harrison (2005)$^{\#}$& 7.89		& 2.5		& 0.458     & 3.36      \\
\tableline
\end{tabular}
\tablecomments
{\\(1) L$>$10$^{12}L_{\odot}$ SED is the model fit for the archetypical dERO HR10\\
(3) Salpeter IMF is assumed\\
(4) $\Theta$ is the opening angle of the star forming region\\
(5) Small Magellanic Cloud extinction curve is assumed\\
($\#$) Converted from H$_{0}=65$ kms$^{-1}$Mpc$^{-1}$ in Serjeant \& Harrison (2005)  to H$_{0}=70$ kms$^{-1}$Mpc$^{-1}$}\\
\end{center}
\end{table}

\begin{table}
\begin{center}
\caption{The parameters for the evolutionary models for dEROs described in the text
\label{table:dEROevolution}}
\begin{tabular}{ c c c c c}
\tableline \tableline
 Population 	& \multicolumn{2}{c}{evolution to z=1} 	& \multicolumn{2}{c}{evolution to z=3} \\
		    	&     $k_{1}$		&     $g_{1}$        		& $k_{2}$          &  $g_{2}$        \\
  (1)   	    	& (2)			& (3)   				& (4)                  & (5)                 \\
 \tableline
STFG		& 3.2			& 0.9				& 2.2                  &  0.4                \\
LIRG		& 2.5			& 3.2				& 2.2                  &  2.7                \\
ULIRG		& 2.7			& 3.3				& 2.2                  &  2.9                 \\
\tableline
\end{tabular}
\end{center}
\end{table}

\end{document}